\documentclass{vgtc}                          

\graphicspath{{figures/}{pictures/}{images/}{./}} 

\usepackage{times}                     

\usepackage{mathptmx}                  
\usepackage[svgnames]{xcolor}

\onlineid{0}

\vgtccategory{Research}

\vgtcinsertpkg

\title{Using Real Names of Disabled Participant-Contributors to Practice Citational Justice in Accessibility}

\author{%
Jonathan Zong\\ %
    \scriptsize University of Colorado Boulder%
}

\abstract{
In accessibility research involving human subjects, researchers conventionally anonymize their research participants to protect privacy.
However, a lack of intentionality about who to publicly acknowledge for intellectual contributions to research can lead to the erasure of disabled individuals’ work and knowledge.
In this paper, I propose identifying disabled research participants by name (with consent) as a practice of citational justice.
I share observations from examples of this practice in accessible visualization research, and offer considerations for when it may be appropriate to de-anonymize.
Intentional practices of citation offer researchers an opportunity to acknowledge the expertise and intellectual contributions of disabled people in our communities.
} 


\begin{document}
\firstsection{Introduction}
\maketitle

In accessibility research involving human subjects, researchers conventionally anonymize their research participants to protect privacy through the use of pseudonyms or coded identifiers like ``Participant 1.''
This practice emerged from well-intentioned motivations: to shield participants from potential harm and to maintain the perceived objectivity of research findings by distancing individual voices from collective insights.

However, the standard practice of anonymization can inadvertently contribute to the systematic erasure of disabled individuals' intellectual work and expertise.
When disabled people participate in accessibility research, they bring multiple forms of specialized knowledge: lived experience of disability, deep technical expertise with assistive technologies, and professional insights from navigating inaccessible systems.
Yet when these contributions are anonymized, the scholarly record fails to acknowledge many individuals whose ideas and innovations drive progress in the field.
Given that accessibility research relies heavily on the expertise of disabled people, it is important that we not only credit researchers, but also community members whose knowledge makes research possible.

This paper proposes that identifying disabled research participants by name---with their informed consent---can serve as a practice of citational justice in accessibility research.
Rather than abandoning anonymization entirely, I argue for intentional decisions about when de-anonymization is appropriate and beneficial.
Drawing from examples in accessible visualization research with blind and low-vision participants and co-designers, I demonstrate how named attribution can honor participants' expertise while maintaining ethical research practices.

\subsection{Positionality Statement}

I am an accessible visualization researcher who works with blind and low-vision (BLV) people to design interfaces for non-visual data exploration.
I am also a research ethics scholar who has studied consent in online research.
As a researcher without lived experience of blindness, my work with BLV people has required me to be attentive to the role of different forms of expertise.
Because of the position that I occupy as a university researcher and the role that I play in the type of research I do, my perspective has limitations.
However, I intend this paper as a contribution to an ongoing interdisciplinary conversation about context-specific ways to think about ethics practices.

\section{Background}

\subsection{Anonymity and Pseudonyms in Research Ethics}
\label{sec:background}

Why is it standard practice for researchers to hide the identities of study participants? Anthropologist Carole McGranahan notes two common reasons: \textit{addressing risk} and \textit{claiming truth} \cite{carole_mcgranahan_truths_2021}.

From the perspective of addressing risk, anonymity offers research participants privacy and plausible deniability.
This can be important especially for people who can be put at risk of harassment or violence when they share sensitive information.
In some cases, participants might not want to speak candidly unless they are offered privacy.

From the perspective of claiming truth, the use of pseudonyms serves to bridge the gap between individual and collective experience.
Qualitative data is grounded in individual experience, but researchers want to produce insights that ``hold possibly true for anyone in a collective'' \cite{carole_mcgranahan_truths_2021}.
Distancing an idea from an individual identity reinforces the idea that a participant represents a broader collective.

But recently, researchers (particularly in anthropology) have come to reflect critically on these norms and their limitations.
Weiss and McGranahan point out that the need for pseudonyms is taken for granted such that researchers rarely reflect intentionally on what purpose they are serving \cite{erica_weiss_rethinking_2021}.
There can be many reasons why anonymization and pseudonyms may not be appropriate for a given research context.
For example, scholars have argued that using real names makes researchers more ``acutely and transparently responsible for what we say about people'' and leave the potential for possible ``conversations or rebuttals'' \cite{erica_weiss_rethinking_2021}.
And as I discuss in the next section, an important ethical dimension has to do with properly citing and valuing knowledge from people who may not only be research participants, but also colleagues and intellectual partners.
These critical conversations suggest that researchers should make intentional decisions about how to identify and represent their interlocutors.

\subsection{Citational Justice in HCI and Disability}
The pursuit of \textit{citational justice} is an intentional practice of acknowledging and citing intellectual contributions to address systemic biases in academic citation.
As part of a growing citational justice movement in HCI \cite{kishonna_gray_citeherwork_2015, collective_following_2021, the_citational_justice_collective_citational_2022}, scholars have pointed out that citation can reflect unjust values about which people are ``worthy of being listened to,'' what kinds of contributions ``count as credible'' or are forms of ``valued knowledge'' in HCI \cite{the_citational_justice_collective_citational_2022, sum_challenging_2024}.
This has led to a ``call for collective action'' \cite{kumar_braving_2021} in the community to critically examine citation practices and reimagine them to account for pluralistic modes of knowledge production.
It is against this context that the use of pseudonyms can be seen as an anti-citational practice, one which hinders researchers' ability to ``engage in relationships of meaningful reciprocity with [their] interlocutors'' \cite{erica_weiss_pseudonyms_2021}.

Citational justice also closely intersects with conversations about authorship, and whether the mechanism of authorship leads communities to value certain kinds of labor or contribution at the expense of others \cite{the_citational_justice_collective_citational_2022}.
In conversations on \textit{crip authorship}, scholars have pointed out that there are usually many people who make research possible other than named authors.
Further, when researchers identify others as ``research subjects,'' they hold power to define and represent those people in ways that may conflict with how they understand themselves \cite{mills_introduction_2023}.
This strict dichotomy between researchers and subjects has been especially harmful in the context of disability research that excludes the perspectives of disabled people, which researchers have named as a form of \textit{epistemic violence} \cite{ymous_i_2020}.
These are questions about how knowledge about disabled people can come from sources that do not fit within the conventional researcher/participant dichotomy that conventional authorship reflects.

\section{Enacting Citational Justice in Accessibility}

In this section, I first make the argument for citing disabled participants by name in accessibility research. Next, I discuss potential limitations of this approach. I then offer practical considerations for researchers who are considering doing this in their own work. Finally, I discuss examples and reflections from existing accessible visualization research.

\subsection{Citing Participants by Name}

In HCI accessibility research, it's common to expect researchers to involve disabled people in research. 
When disabled people are part of a project's core research team, they usually receive credit as an author of research papers associated with that project.
Authorship here signifies and credits an intellectual contribution to the research.
But when disabled people are sources of qualitative data for research, as participants in user studies or interviews, it is much less common that they receive direct credit for these contributions.
Even though they are contributing original ideas as part of the research, and the research cannot proceed without them, people who are categorized as research subjects are usually anonymized.

In this section, I argue that crediting disabled participants by name (with consent) can be a form of citational justice.
To be clear, I am not advocating that we completely abandon anonymization as an ethics procedure.
The norm to use anonymization was created for good reason, to protect participants from harm that can arise from increased attention and prevent abuse of power by researchers.
However, it's important that researchers make intentional decisions about naming, so that we do not miss opportunities to treat participants and their ideas with respect.

Accessibility research has highly specific requirements for participant expertise.
For example, participants in my own research on accessible visualization have multiple forms of expertise: they are screen reader experts, work with data in their everyday lives, and have lived experience of blindness.
Building relationships with these participants requires working in partnership with existing communities and groups, especially in my local area.
Because of these specific needs, the community and participant pool is relatively small compared to other kinds of research.
As a result, I'm often relying on the same people across multiple studies --- cultivating long-term relationships.

These experiences demonstrate that beyond who we typically consider a researcher, there is a broader set of people who can consistently contribute to the research conversation behind the scenes.
As disability scholars have pointed out, the use of pseudonyms to anonymize these people comes with the ``assumption that there is a researcher who has the `theory' and the `method' and there are participants who do not'' \cite{edwards_going_2024}.
But many people who have engaged with my research process are trailblazers who act incredibly resourcefully to develop their own best practices and ways of working.
For example, some participants are busy working professionals who see contributing to research studies as an act of service to the disability community, and their availability shouldn't be taken for granted.
Unless we publicly ``acknowledge our debt to those who came before,'' \cite{ahmed_living_2017} we risk ``withhold[ing]
important historical information that may not be available elsewhere'' about where research discoveries in accessibility come from \cite{edwards_going_2024}. 

\subsection{Limitations of Citation}

For academic researchers, citation has clear value as a form of reputation and credibility --- but do citations carry any value for non-academics who are part of research?
For example, people have a variety of reasons for participating in accessibility research studies.
As my colleague Daniel Hajas points out, ``Perhaps they are curious about a particular research prototype. Perhaps, they need the monetary compensation and jump from one study to another to make an income. There may be several motivators, and we can't always assume the ideal and altruistic nature of participation.''
Latour and Woolgar have referred to credibility as a form of capital, which ``makes possible the conversion between money, data, prestige,
credentials, problem areas, argument, papers, and so on'' in an endless cycle of investment \cite{bruno_latour_cycle_1982}.
But for non-academics who are not part of this cycle, it can be less clear how they benefit from recognition.

Furthermore, citational justice scholars have pointed out ways that citation can sometimes do harm.
When quotes or arguments are taken out of context, they can be misrepresented and used to support claims that were not intended by the original speaker \cite{the_citational_justice_collective_citational_2022}.
In other cases, attributing statements to individuals when they are speaking in a professional context on behalf of an organization can expose that person to career repercussions \cite{miia_halme-tuomisaari_between_2021}.
And in the case of paper authorship, granting authorship inappropriately can lead to harm.
For example, it is harmful when disabled co-authors do not have the power or ability to meaningfully contribute to the work, but are tokenized \cite{liang_embracing_2021} by non-disabled co-authors to benefit from the perceived credibility of having a disabled co-author.

But on the other hand, there are also cases where credit for intellectual work does have value outside of academia.
Weiss gives examples where ethnographic subjects have said that ``if [she] was speaking about them explicitly they would invite all of their supporters, or [if she] wrote about them directly, they could send [her] writing to would-be donors to demonstrate their active engagement with academia'' \cite{erica_weiss_pseudonyms_2021}.
In the HCI accessibility community, disabled researchers who lack formal academic positions or institutional support might find it even more important that academic researchers cite their work.
For example, non-academic disability scholar Liz Jackson has advocated for citational justice, as credit for ideas can be helpful for grants or book deals that would enable future intellectual work that happens outside of a university \cite{jackson_disability_2022}.

As academics, we should not make assumptions about how people will feel about being credited by name for their ideas.
Some will find this valuable and others will not.
Instead, I suggest that we are guided by the value of \textit{reciprocity}.
When people contribute to our research, especially in the context of continued engagement over time, we are receiving their help in the form of time and expertise.
Considering others' specific goals and motivations, it can be helpful to consider what the best way is to give back.
For example, monetary compensation for disabled study participants is a best practice \cite{lundgard_sociotechnical_2019} and should be considered whether or not participants are also named.
In other cases that do not fit into the researcher / participant dichotomy, there are other forms of engagement: including people as part of narrative synthesis in evidence reviews, naming people in a paper's acknowledgments, and co-authorship.

\subsection{Considerations for Ethical Citation}

Given that the practice I am proposing goes against what are seen as conventional research ethics procedures, how should researchers know when to de-anonymize?
Here, I offer considerations to researchers who are exploring whether they should reference participants by name in their work.
Decisions relevant to research ethics should be made as part of a standard university Institutional Review Board (IRB) process.

\begin{itemize}

    \item \textbf{Ensure minimal risk and get IRB approval.}
Consider the typical IRB criteria for minimal risk social and behavioral research.
De-anonymizing would not be appropriate for research that is not minimal risk.
Think about any potential risks (especially social or reputational) to participants that may arise from being associated with their quotes.
Some researchers are unaware that IRBs do allow the use of real names in certain cases \cite{erica_weiss_rethinking_2021}.
Include the intention to use real names in IRB materials so that ethics reviewers can help determine whether it is appropriate to proceed.

    \item \textbf{Ask for consent.}
Ask each participant for informed consent to use their name.
Explain the context in which quotes will be used, and explain the audience who will be reading the research paper.
Consider giving a granular set of options for how they would like to be quoted.
For example, this can include the use of a pseudonym, referring to them by their relevant identity or job role (e.g. blind assistive technology expert), using just their first name, or using their full name.
    
    \item \textbf{Value reciprocity.}
Remember that the point of using people's names is to attribute ideas and lift them up for their contributions.
Consider having a broader conversation about how contributing to research might advance their goals.
These conversations can be an important part of communicating expectations \cite{lundgard_sociotechnical_2019} about research process.
If participants do not feel that citing them by name productively contributes toward their goals, then explore alternatives.
    
\end{itemize}

\subsection{Examples from Accessible Visualization Research}

In this section, I share a few examples from my own work in accessible data visualization that I consider success stories for citing contributors by name. \\

\subsubsection{Quoting a co-author/co-designer by name in writing.}

In the \textit{Rich Screen Reader Experiences for Accessible Data Visualization} \cite{zong_rich_2022} project, we developed a set of design dimensions for screen reader interfaces.
To develop these design dimensions, authors with data visualization and interface design expertise engaged in an iterative co-design process with blind co-author Daniel Hajas.
Daniel brought several forms of expertise to the project, including lived experience of blindness and research experience designing assistive technology.
In our co-design process, we would conduct weekly meetings on Zoom and exchanged lengthy emails that included Daniel's reflections on iterations of our design prototypes.

Daniel was a co-author of the paper, but we decided to also include quotes from his written notes and correspondence in order to share insights from our co-design process.
We attributed these quotes to him by name.
This went against conventional notions that authors of research papers should speak with a single collective voice, and challenged conventional separations between author and co-designer.
Though reviewers initially raised questions about this choice, we successfully by addressed these concerns by expanding our methods section with notes that clarified author positionality. \\

\subsubsection{De-anonymizing BLV experts in a user study.}

In \textit{Umwelt: Accessible Structured Editing of Multi-Modal Data Representations} \cite{zong_umwelt_2024}, we designed and evaluated an accessible editor for BLV users to create data representations that include visualization, sonification, and structured textual descriptions.
Because our system was designed with expert users in mind, the user study had fairly restrictive eligibility criteria.
As such, we opted to do in-depth evaluations (3 hours total per participant) with a smaller number of people.

All five participants had high levels of experience with assistive technology, working with data, or both.
Most of them were people who I had met through previous research, and were consistently valuable contributors to the work.
In the paper, I quote from highly situated stories they share about their own workflows and professional contexts.
With permission from my IRB, I individually asked each participant for consent to use their name with their quotes in the paper, and all five gave consent.

\subsubsection{Clarifying author contributions to co-design}

In \textit{Data Navigator: An Accessibility-Centered Data Navigation Toolkit} \cite{elavsky_data_2023}, Elavsky et al. created a toolkit for developers to create keyboard navigable structures for screen reader users.
In that paper, the authors explicitly stated what contributions were made by blind co-designer Lucas Nadolskis, who was also a co-author of the paper.
According to first author Frank Elavsky, ``Lucas and I had discussions about including this, and we felt it was not only informative for others who might do co-design down the road, but also good for the citational justice aspect of the work --- similarly to how the Contributor Role Taxonomy (CRediT) \cite{holcombe_contributorship_2019} works, making contributions as transparent as possible.''
For this team, it was important that even though Lucas was included as a co-author, it was clear what level of involvement and agency he had over the work.
Elavsky says, ``Citational justice can also be used as a way to make a paper seem more authoritative than it really is. I think a danger of including people in a citation is if a bad actor does it (someone who didn't actually give participants real authorial/intellectual freedom, yet wants to benefit from having a disabled co-author). I wanted to be really transparent that Lucas's involvement was relatively small and to be clear about when he had decision-making power and agency.''
This way, the team balanced publicly assigning credit with being clear about the scope of everyone's role and contributions.

\section{Conclusion}

While pseudonyms arose from an important set of research ethics considerations, I argue that they are a solution that does not fit the context of low-risk HCI accessibility research.
First, participants are usually not placed at risk for being associated with their knowledge about accessibility technology.
Second, the idea that knowledge should attempt to be generalizable across people and time is a poor match for disability justice values, which acknowledge that there are no normative bodies or one-size-fits-all solutions \cite{sum_challenging_2024, hamraie_designing_2013}.
Connecting ideas with specific identities and situations is a way of acknowledging that accessibility innovations are born out of situated experiences and knowledges \cite{haraway_situated_1988}.
In this paper, I suggest that the HCI community consider intentional practices of citation in accessibility research.
By crediting participants by name when appropriate, we can build a more just scholarly record and cultivate reciprocal relationships between academics and our adjacent communities.

\acknowledgments{
Thanks to Frank Elavsky for encouraging me to write these thoughts down. Frank Elavsky, Daniel Hajas, and Crystal Lee gave feedback on drafts of this paper. Thank you!}

\bibliographystyle{abbrv-doi}

\bibliography{citational-justice}
\end{document}